\documentclass[final,5p,times,twocolumn]{elsarticle}
\usepackage{epsfig}
\usepackage{amsmath,amssymb,amsfonts,slashed}
\usepackage{xcolor}
\usepackage{bm}
\usepackage{booktabs}
\usepackage{graphicx}
\usepackage[
    colorlinks=true,
    linkcolor=blue,
    citecolor=blue,
    urlcolor=blue
]{hyperref}

\journal{Physics Letters B}
\biboptions{sort&compress}
\begin{document}

\begin{frontmatter}

\title{On-shell constituent approach to the light-front partonic energy-momentum tensor}

\author[1]{Abi Jebarson A}
\ead{abijebarson@gmail.com}
\author[1]{Harleen Dahiya}
\ead{dahiyah@nitj.ac.in}
\affiliation[1]{organization={Department of Physics, Dr. B R Ambedkar National Institute of Technology},
            city={Jalandhar},
            state={Punjab},
            postcode={144008},
            country={India}}

\begin{abstract}
    The forward matrix elements of the energy-momentum tensor in momentum space, parameterized by gravitational transverse-momentum distributions (TMDs), provide a microscopic window into hadronic mechanical structure. While the longitudinal and transverse columns match to canonical vector TMDs up to twist-4, the light-front minus column involves the unintegrated parton energy $k^-$, leaving this sector undetermined in general quantum chromodynamics. Working within relativistic light-front constituent frameworks where active quarks are modeled as effective on-shell quasiparticles, we show that this energy factorizes algebraically. This establishes a closed matching relating all 32 gravitational TMDs to canonical vector TMDs up to twist-4, including target polarization and T-odd final-state interactions. This completion provides access to the light-front energy density and its transverse Sivers dipole distortion, demonstrates the local equivalence between convective energy and momentum transport velocities, and determines the longitudinal normal stress. Comparing longitudinal and transverse stresses yields a boost-invariant anisotropy ratio that charts the mechanical crossover from collinear confinement dominance to transverse isotropic balance. While formulated within the constituent approximation, these results offer baseline insights into the three-dimensional continuum mechanics of the nucleon.
\end{abstract}

\end{frontmatter}

\section{Introduction}
\label{sec:intro}

The energy-momentum tensor (EMT) of quantum chromodynamics (QCD) provides a fundamental framework for characterizing the internal structure of hadrons~\cite{Ji:1994av,Pagels:1966zza,Kobzarev:1962wt}. It is directly related to the longstanding questions of the hadron mass~\cite{Ji:1994av,Ji:1995sv,Yang:2018nqn,Lorce:2017xzd,Metz:2020vxd} and spin decompositions~\cite{Jaffe:1989jz,Ji:1996ek,Leader:2013jra,Wakamatsu:2014zza}, and allows one to investigate internal mechanical properties such as pressure, shear forces, and mechanical stability~\cite{Polyakov:2002yz,Burkert:2018bqq,Lorce:2018egm,Polyakov:2018zvc,Freese:2021czn}. Probing the matrix elements of the EMT is therefore a central objective of the physics program at the future Electron-Ion Collider (EIC)~\cite{Burkert:2023wzr,AbdulKhalek:2021gbh,AbdulKhalek:2022hcn}.

Although direct microscopic scattering via gravitational probes is experimentally unfeasible, the local EMT in QCD can be probed indirectly through hard exclusive processes parameterized by gravitational form factors~\cite{Kobzarev:1962wt,Pagels:1966zza,Bakker:2004ib}, which in turn connect to generalized parton distributions (GPDs)~\cite{Ji:1996ek,Diehl:2003ny,Lorce:2015lna,Guo:2021aik} and generalized distribution amplitudes (GDAs)~\cite{Kumano:2017lhr} in coordinate and impact-parameter space. While the connection between the EMT and GPDs is well established, the link with transverse-momentum-dependent distributions (TMDs) in momentum space has only recently been formulated through the introduction of gravitational TMDs~\cite{Lorce:2023zzg}, alongside related investigations of internal forces and higher-twist structures~\cite{Burkardt:2003yg,Boer:2015vso,Amor-Quiroz:2020qmw,Angeles-Martinez:2015sea,Kaur:2026wxm,Kaur:2026jjr}. For a spin-$\frac{1}{2}$ hadron, 32 independent gravitational TMDs emerge. As shown in Ref.~\cite{Lorce:2023zzg}, the $\nu = +$ and $\nu = i$ components of the momentum-space EMT match directly to canonical vector TMDs up to twist-4. In contrast, the light-front minus column ($\nu = -$) involves the unintegrated quark light-front energy $k^-$, leaving this sector undetermined in general QCD.

The aim of the present work is to explore this light-front minus sector within relativistic light-front constituent frameworks~\cite{Brodsky:1994fz,Schlumpf:1992pp,Pasquini:2008ax,Lorce:2011kd}. By treating active quarks as effective on-shell quasiparticles, the light-front energy factorizes algebraically, allowing all 32 gravitational TMDs to be determined in terms of canonical vector TMDs up to twist-4. This formulation provides a transparent, boost-invariant quasiparticle baseline for the three-dimensional (3D) continuum mechanics of the hadron in momentum space, including the partonic light-front energy density, convective transport velocities, and the longitudinal-to-transverse stress anisotropy.

We organize this paper by formulating the on-shell kinematics and matching relations connecting gravitational TMDs to canonical vector TMDs in Section~\ref{sec:matching}. In Section~\ref{sec:mechanics}, we analyze the resulting momentum-space mechanical properties. We illustrate these distributions within a relativistic quark-diquark framework in Section~\ref{sec:numerical}, and present our conclusions in Section~\ref{sec:conclusion}.

\section{On-shell kinematics and matching relations}
\label{sec:matching}

\begin{table*}[t]
    \caption{Matching of all 32 gravitational TMDs for a spin-$\frac{1}{2}$ hadron in terms of canonical vector TMDs up to twist-4 under the light-front on-shell condition, with $\mathcal{K} = \frac{m^2+\boldsymbol{k}_\perp^2}{2xM^2} - \frac{x}{2}$, following the tensor classification of Ref.~\cite{Lorce:2023zzg}.}
    \label{tab:all_gravitational_TMDs}
    \centering
    \setlength{\tabcolsep}{6pt}
    \renewcommand{\arraystretch}{1.15}
    \begin{tabular}{cccc|ccc}
        \hline\hline
        $j$ & Tensor Basis       & $a_j$ (Unpolarized)                  & $a_{jT}^\perp$ (T-odd)             & Pseudotensor Basis               & $a_{jL}$ (Longitudinal) & $a_{jT}$ (Transverse) \\
        \hline
        $0$ & $M^2 g_T^{\mu\nu}$ & $0$                               & $0$                                                 & -                                & -                       & -                     \\
        $1$ & $P^\mu P^\nu$      & $x f_1$                           & $x f_{1T}^\perp$                                    & $P^\mu \epsilon_T^{\nu \cdot}$   & $0$                     & $0$                   \\
        $2$ & $N^\mu N^\nu$      & $\mathcal{K}(f_3 - \tfrac{1}{2}f_1)$ & $\mathcal{K}(f_{3T}^\perp - \tfrac{1}{2}f_{1T}^\perp)$ & $P^\nu \epsilon_T^{\mu \cdot}$   & $x f_L^\perp$           & $x f_T^+$             \\
        $3$ & $k_T^\mu k_T^\nu$  & $f^\perp$                         & $f_T^\perp$                                         & $N^\mu \epsilon_T^{\nu \cdot}$   & $0$                     & $0$                   \\
        $4$ & $P^\mu N^\nu$      & $\mathcal{K} f_1$                    & $\mathcal{K} f_{1T}^\perp$                             & $N^\nu \epsilon_T^{\mu \cdot}$   & $\mathcal{K} f_L^\perp$    & $\mathcal{K} f_T^+$      \\
        $5$ & $N^\mu P^\nu$      & $x (f_3 - \tfrac{1}{2}f_1)$       & $x (f_{3T}^\perp - \tfrac{1}{2}f_{1T}^\perp)$       & $k_T^\mu \epsilon_T^{\nu \cdot}$ & $0$                     & $0$                   \\
        $6$ & $P^\mu k_T^\nu$    & $f_1$                             & $f_{1T}^\perp$                                      & $k_T^\nu \epsilon_T^{\mu \cdot}$ & $f_L^\perp$             & $f_T^+$               \\
        $7$ & $k_T^\mu P^\nu$    & $x f^\perp$                       & $x f_T^\perp$                                       & -                                & -                       & -                     \\
        $8$ & $N^\mu k_T^\nu$    & $f_3 - \tfrac{1}{2}f_1$           & $f_{3T}^\perp - \tfrac{1}{2}f_{1T}^\perp$           & -                                & -                       & -                     \\
        $9$ & $k_T^\mu N^\nu$    & $\mathcal{K} f^\perp$                & $\mathcal{K} f_T^\perp$                                & -                                & -                       & -                     \\
        \hline\hline
    \end{tabular}
\end{table*}

We adopt standard light-front conventions $v^\mu = [v^+, v^-, \boldsymbol{v}_\perp]$ with $v^\pm = (v^0 \pm v^3)/\sqrt{2}$, metric components $g_{+-} = g_{-+} = 1$, $g_{ij} = -\delta_{ij}$ ($i,j \in \{1,2\}$), and scalar product $a \cdot b = a^+ b^- + a^- b^+ - \boldsymbol{a}_\perp \cdot \boldsymbol{b}_\perp$~\cite{RevModPhys.21.392,Brodsky:1997de}. For a hadron of mass $M$ and four-momentum $P^\mu = [P^+, M^2/(2P^+), \boldsymbol{0}_\perp]$, the symmetric projector onto the transverse subspace is $g_T^{\mu\nu} = g^{\mu\nu} - (P^\mu N^\nu + P^\nu N^\mu)/M^2 + N^\mu N^\nu/M^2$, where $N^\mu = [0, M^2/P^+, \boldsymbol{0}_\perp]$ is the lightlike vector directed along the Wilson line ($N^2 = 0, P \cdot N = M^2$). Any four-vector expands in the basis $\{P^\mu, N^\mu, a_T^\mu\}$ as
\begin{equation}
    a^\mu = \frac{a^+}{P^+}\left( P^\mu - \frac{1}{2}N^\mu \right) + \frac{a^- P^+}{M^2} N^\mu + a_T^\mu.
    \label{eq:vector_decomp}
\end{equation}

In relativistic light-front constituent models, the active quark carries an effective quasiparticle mass $m$. Setting the quark on its light-front mass shell, $k^2 = 2k^+ k^- - \boldsymbol{k}_\perp^2 = m^2$, fixes the light-front energy algebraically as
\begin{equation}
    k^- = \frac{m^2 + \boldsymbol{k}_\perp^2}{2xP^+}.
    \label{eq:kminus_onshell}
\end{equation}
Substituting Eq.~\eqref{eq:kminus_onshell} into the quark four-momentum $k^\mu = [x P^+, k^-, \boldsymbol{k}_\perp]$ and expanding it via Eq.~\eqref{eq:vector_decomp} yields
\begin{equation}
    k^\mu = x P^\mu + k_T^\mu + \mathcal{K}(x, \boldsymbol{k}_\perp^2) N^\mu,
    \label{eq:k_expansion}
\end{equation}
where $x = k^+/P^+$, $k_T^\mu = [0, 0, \boldsymbol{k}_\perp]$, and the dimensionless kinematic function is
\begin{equation}
    \mathcal{K}(x, \boldsymbol{k}_\perp^2) \equiv \check{\mathcal{K}}(x, \boldsymbol{k}_\perp^2) - \frac{x}{2} = \frac{m^2 + \boldsymbol{k}_\perp^2}{2xM^2} - \frac{x}{2}.
    \label{eq:K_def}
\end{equation}

The quark vector TMD correlator $\Phi^{[\gamma^\mu]}$ for a spin-$\frac{1}{2}$ hadron is parameterized up to twist-4 by eight canonical TMDs~\cite{Sivers:1989cc,Mulders:1995dh,Boer:1997nt,Goeke:2005hb,Bacchetta:2006tn,Rodini:2022wki,Kaur:2026jjr}, where $f_1, f_{1T}^\perp$ are twist-2, $f^\perp, f_T, f_T^\perp, f_L^\perp$ are twist-3, and $f_3, f_{3T}^\perp$ are twist-4,
\begin{equation}
    \begin{aligned}
        \Phi^{[\gamma^+]}        & = f_1 - \frac{\epsilon_T^{k_T S_T}}{M}\, f_{1T}^\perp,                                                                                                                                   \\
        \Phi^{[\gamma_T^\alpha]} & = \frac{1}{P^+} \Big[ k_T^\alpha \left( f^\perp - \frac{\epsilon_T^{k_T S_T}}{M}\, f_T^\perp \right) - M \epsilon_T^{\alpha S_T} f_T^+ - \lambda \epsilon_T^{\alpha k_T} f_L^\perp \Big], \\
        \Phi^{[\gamma^-]}        & = \left(\frac{M}{P^+}\right)^{\!2} \left[ f_3 - \frac{\epsilon_T^{k_T S_T}}{M}\, f_{3T}^\perp \right],
        \label{eq:phi_components}
    \end{aligned}
\end{equation}
where $\epsilon_T^{\mu\nu} = \epsilon^{\mu\nu\alpha\beta} N_\alpha P_\beta / M^2$ (with $\epsilon_T^{12} = +1$). For any transverse vectors $a_T$ and $b_T$, we use the compact notations $\epsilon_T^{\mu a_T} \equiv \epsilon_T^{\mu\nu} a_{T\nu}$ and $\epsilon_T^{a_T b_T} \equiv \epsilon_T^{\alpha\beta} a_{T\alpha} b_{T\beta} = (\boldsymbol{a}_\perp \times \boldsymbol{b}_\perp)_z$, such that $\epsilon_T^{k_T S_T} = (\boldsymbol{k}_\perp \times \boldsymbol{S}_\perp)_z$. Furthermore, $\lambda$ is hadron helicity, $S_T^\mu = [0, 0, \boldsymbol{S}_\perp]$ is transverse spin, and $f_T^\pm \equiv f_T \pm \frac{\boldsymbol{k}_\perp^2}{2M^2} f_T^\perp$. Reconstructing the full four-vector yields
\begin{equation}
    \begin{aligned}
        \Phi^{[\gamma^\mu]} = \frac{1}{P^+} \Bigg\{ & P^\mu \left( f_1 - \frac{\epsilon_T^{k_T S_T}}{M}\, f_{1T}^\perp \right) \\
                                                    & + k_T^\mu \left( f^\perp - \frac{\epsilon_T^{k_T S_T}}{M}\, f_T^\perp \right) \\
                                                    & + N^\mu \left[ \left( f_3 - \frac{1}{2} f_1 \right) - \frac{\epsilon_T^{k_T S_T}}{M} \left( f_{3T}^\perp - \frac{1}{2} f_{1T}^\perp \right) \right]    \\
                                                    & - M \epsilon_T^{\mu S_T} f_T^+ - \lambda \epsilon_T^{\mu k_T} f_L^\perp \Bigg\}.
        \label{eq:phi_fourvector}
    \end{aligned}
\end{equation}

The unintegrated quark EMT correlator is $\Theta_q^{\mu\nu}(P, k) = k^\nu W^{[\gamma^\mu]}(P, k)$, where $W^{[\gamma^\mu]}(P, k)$ is the unintegrated quark-quark correlator~\cite{Meissner:2009ww} associated with the gauge-invariant canonical (gic) quark EMT operator $T_{q,\text{gic}}^{\mu\nu}(z) = \bar{\psi}(z) \gamma^\mu \frac{i}{2} \overleftrightarrow{D}_{\text{pure}}^\nu \psi(z)$~\cite{Lorce:2023zzg,Leader:2013jra} with a future- or past-pointing Wilson line. In light-front gauge $A^+=0$, the pure-gauge covariant derivative $D_{\text{pure}}^\nu$ reduces to the ordinary derivative $\partial^\nu$ conjugate to the canonical momentum $k^\nu$. In general QCD, integrating $\Theta_q^{\mu\nu}(P, k)$ over $k^-$ factorizes directly only for the $\nu \neq -$ components, $\mathcal{T}_q^{\mu\nu} = \Phi^{[\gamma^\mu]} k^\nu$ ($\nu \neq -$)~\cite{Lorce:2023zzg}, leaving the light-front minus column undetermined. Within the present constituent framework, however, the light-front energy is algebraically fixed by the on-shell condition via Eq.~\eqref{eq:kminus_onshell}. Consequently, the factorization extends directly to all components, yielding
\begin{equation}
    \mathcal{T}_q^{\mu\nu}(P, x, \boldsymbol{k}_\perp, S) = \Phi^{[\gamma^\mu]}(P, x, \boldsymbol{k}_\perp, S)\, k^\nu.
    \label{eq:T_factorization}
\end{equation}
Inserting Eq.~\eqref{eq:k_expansion} into Eq.~\eqref{eq:T_factorization}, the TMD EMT decomposes as
\begin{equation}
    \mathcal{T}_q^{\mu\nu} = x \Phi^{[\gamma^\mu]} P^\nu + \Phi^{[\gamma^\mu]} k_T^\nu + \mathcal{K}(x, \boldsymbol{k}_\perp^2)\, \Phi^{[\gamma^\mu]} N^\nu.
    \label{eq:T_decomposed}
\end{equation}

Matching Eq.~\eqref{eq:T_decomposed} onto the model-independent parametrization of the TMD EMT~\cite{Lorce:2023zzg} determines all 32 gravitational TMDs within this framework, summarized in Table~\ref{tab:all_gravitational_TMDs}. In this classification, the unpolarized ($a_j$) and transverse T-odd ($a_{jT}^\perp$) sectors are expanded across the ten basis tensors ($j=0,\dots,9$), while the longitudinal ($a_{jL}$) and transverse ($a_{jT}$) sectors are expanded across the six basis pseudotensors $P^\mu \epsilon_T^{\nu \cdot}$, $P^\nu \epsilon_T^{\mu \cdot}$, $N^\mu \epsilon_T^{\nu \cdot}$, $N^\nu \epsilon_T^{\mu \cdot}$, $k_T^\mu \epsilon_T^{\nu \cdot}$, and $k_T^\nu \epsilon_T^{\mu \cdot}$ ($j=1,\dots,6$). The 18 relations for $\nu \in \{+, i\}$ derived in Ref.~\cite{Lorce:2023zzg} are directly recovered. For the unintegrated $\nu = -$ functions $\check{F} \in \{\check{f}_1, \check{f}_{1T}^\perp, \check{f}^\perp, \check{f}_T^\perp, \check{f}_3, \check{f}_{3T}^\perp, \check{f}_L^\perp, \check{f}_T^+\}$, we obtain the direct proportionality as
\begin{equation}
    \check{F}(x, \boldsymbol{k}_\perp^2) = \frac{m^2 + \boldsymbol{k}_\perp^2}{2xM^2}\, F(x, \boldsymbol{k}_\perp^2) = \check{\mathcal{K}}(x, \boldsymbol{k}_\perp^2) F(x, \boldsymbol{k}_\perp^2).
    \label{eq:constituent_check}
\end{equation}
Eq.~\eqref{eq:constituent_check} provides the central matching relation of this constituent approach, algebraically determining all $\nu = -$ gravitational TMDs directly from their canonical vector TMD partners.

\section{Light-front EMT components and mechanical properties}
\label{sec:mechanics}

Within this constituent description, all nine light-front components $\mathcal{T}_q^{\mu\nu}$ ($\mu, \nu \in \{+, i, -\}$) are determined with full target polarization $(\lambda, \boldsymbol{S}_\perp)$ as
\begin{subequations}
    \label{eq:all_components}
    \begin{align}
        \mathcal{T}_q^{++} & = x P^+ \left( f_1 - \frac{\epsilon_T^{k_T S_T}}{M}\, f_{1T}^\perp \right), \label{eq:Tpp}                                                                                                                                              \\
        \mathcal{T}_q^{+i} & = k_T^i \left( f_1 - \frac{\epsilon_T^{k_T S_T}}{M}\, f_{1T}^\perp \right), \label{eq:Tpi}                                                                                                                                              \\
        \mathcal{T}_q^{+-} & = \left( \frac{m^2 + \boldsymbol{k}_\perp^2}{2x P^+} \right) \left( f_1 - \frac{\epsilon_T^{k_T S_T}}{M}\, f_{1T}^\perp \right), \label{eq:Tpm}                                                                                               \\
        \mathcal{T}_q^{i+} & = x \Big[ \left( f^\perp - \frac{\epsilon_T^{k_T S_T}}{M}\, f_T^\perp \right) k_T^i \notag \\
                           & \quad\quad - M \epsilon_T^{i S_T} f_T^+ - \lambda \epsilon_T^{i k_T} f_L^\perp \Big], \label{eq:Tip} \\
        \mathcal{T}_q^{ij} & = \frac{1}{P^+} \Big[ \left( f^\perp - \frac{\epsilon_T^{k_T S_T}}{M}\, f_T^\perp \right) k_T^i k_T^j \notag \\
                           & \quad\quad - M \epsilon_T^{i S_T} k_T^j f_T^+ - \lambda \epsilon_T^{i k_T} k_T^j f_L^\perp \Big], \label{eq:Tij} \\
        \mathcal{T}_q^{i-} & = \left( \frac{m^2 + \boldsymbol{k}_\perp^2}{2x (P^+)^2} \right) \Big[ \left( f^\perp - \frac{\epsilon_T^{k_T S_T}}{M}\, f_T^\perp \right) k_T^i \notag \\
                           & \quad\quad - M \epsilon_T^{i S_T} f_T^+ - \lambda \epsilon_T^{i k_T} f_L^\perp \Big], \label{eq:Tim} \\
        \mathcal{T}_q^{-+} & = \frac{M^2}{P^+} x \left( f_3 - \frac{\epsilon_T^{k_T S_T}}{M}\, f_{3T}^\perp \right), \label{eq:Tmp}                                                                                                                                  \\
        \mathcal{T}_q^{-i} & = \frac{M^2}{(P^+)^2} k_T^i \left( f_3 - \frac{\epsilon_T^{k_T S_T}}{M}\, f_{3T}^\perp \right), \label{eq:Tmi}                                                                                                                          \\
        \mathcal{T}_q^{--} & = \frac{M^2}{(P^+)^2} \left( \frac{m^2 + \boldsymbol{k}_\perp^2}{2x P^+} \right) \left( f_3 - \frac{\epsilon_T^{k_T S_T}}{M}\, f_{3T}^\perp \right). \label{eq:Tmm}
    \end{align}
\end{subequations}
Equations~\eqref{eq:all_components} determine the complete set of nine light-front EMT components directly in terms of canonical vector TMDs up to twist-4, mapping the internal mechanical structure of the hadron into parton momentum space $(x, \boldsymbol{k}_\perp)$. The unpolarized distributions $\mathcal{T}_{q,\text{unpol}}^{\mu\nu}$ isolate the unpolarized vector TMDs $f_1, f^\perp, f_3$. For unpolarized nucleons ($\lambda = 0, \boldsymbol{S}_\perp = 0$), the EMT components depend solely on the scalar transverse momentum $k_\perp = |\boldsymbol{k}_\perp|$, and we write $\mathcal{T}_q^{\mu\nu}(x, k_\perp) \equiv \mathcal{T}_{q,\text{unpol}}^{\mu\nu}(x, \boldsymbol{k}_\perp^2)$.

\subsection{Light-front energy density and Sivers dipole distortion}
The component $\mathcal{T}_q^{+-}$ represents the momentum-space density of light-front energy $k^-$. In full QCD, integrating over the complete partonic phase space across all species $i \in \{q, \bar{q}, g\}$ yields the light-front energy sum rule~\cite{Brodsky:1997de,Lorce:2015lna,Lorce:2018egm},
\begin{equation}
    \begin{aligned}
        \sum_i \int \mathrm{d} x\, \mathrm{d}^2 k_\perp\, \mathcal{T}_i^{+-}(x, \boldsymbol{k}_\perp, \boldsymbol{S}_\perp) 
        & = \sum_i \int \mathrm{d} x\, \mathrm{d}^2 k_\perp\, \mathcal{T}_{i,\text{unpol}}^{+-}(x, \boldsymbol{k}_\perp^2) \\
        & = P^- = \frac{M^2}{2P^+}.
    \end{aligned}
\end{equation}
This sum rule holds for any target polarization, as the spin-dependent Sivers term is odd in transverse momentum and vanishes identically upon azimuthal integration $\int \mathrm{d}^2 k_\perp$. In an effective spectator constituent framework, the active quarks contribute a constituent fraction $\langle k^- \rangle_{\text{quarks}}$ to the total hadronic light-front energy, with the accompanying spectator diquark degrees of freedom carrying the remainder, $\langle k^- \rangle_{\text{quarks}} + \langle k^- \rangle_{\text{spectator}} = P^-$.
For a transversely polarized nucleon, Eq.~\eqref{eq:Tpm} determines the spin-dependent energy distribution as
\begin{equation}
    \begin{aligned}
        \mathcal{T}_q^{+-}(x, \boldsymbol{k}_\perp, \boldsymbol{S}_\perp) = & \left( \frac{m^2 + \boldsymbol{k}_\perp^2}{2x P^+} \right) \\
        & \times \left[ f_1(x, \boldsymbol{k}_\perp^2) - \frac{(\boldsymbol{k}_\perp \times \boldsymbol{S}_\perp)_z}{M}\, f_{1T}^\perp(x, \boldsymbol{k}_\perp^2) \right].
    \end{aligned}
    \label{eq:sivers_shift}
\end{equation}
In the context of the quark number density $\rho_q(x, \boldsymbol{k}_\perp, \boldsymbol{S}_\perp) \equiv \Phi^{[\gamma^+]}$, a lateral dipole distortion perpendicular to target spin is well established through Burkardt's chromodynamic lensing mechanism driven by the Sivers effect~\cite{Burkardt:2002ks,Burkardt:2003uw}. Here, having access to $\mathcal{T}_q^{+-}$ shows that this lensing phenomenon directly shifts the \emph{energy density} of the hadron as well. Because each parton carries light-front energy $k^- = (m^2 + \boldsymbol{k}_\perp^2)/(2xP^+)$, the transverse energy dipole shift $\Delta \mathcal{T}_q^{+-} \equiv \mathcal{T}_q^{+-}(x, \boldsymbol{k}_\perp, \boldsymbol{S}_\perp) - \mathcal{T}_{q,\text{unpol}}^{+-}(x, \boldsymbol{k}_\perp^2) = k^- \Delta \rho_q$ acquires an extra factor of $\boldsymbol{k}_\perp^2$ at large $k_\perp$. Consequently, transverse momentum moments of the energy distortion sample higher $k_\perp$-moments of the Sivers function such as $f_{1T}^{\perp(2)}$, placing greater relative weight on the ultraviolet transverse-momentum tails compared to the conventional number density.

\subsection{Convective transport velocities}
The ratio of transverse momentum flux to momentum density defines the convective transport velocity of inertia, $v_\perp^i \equiv \mathcal{T}_q^{i+}/\mathcal{T}_q^{++}$~\cite{Lorce:2023zzg}. Analogously, in relativistic continuum mechanics, the transport velocity of light-front energy is defined by the ratio of transverse energy flux to energy density, $v_{E,\perp}^i \equiv \mathcal{T}_q^{i-}/\mathcal{T}_q^{+-}$. Owing to the factorized on-shell structure $\mathcal{T}_q^{\mu\nu} = \Phi^{[\gamma^\mu]} k^\nu$, the ratio $\mathcal{T}_q^{i\nu}/\mathcal{T}_q^{+\nu}$ is manifestly independent of the transport index $\nu$, yielding
\begin{equation}
    v_{E,\perp}^i(x, \boldsymbol{k}_\perp) \equiv \frac{\mathcal{T}_q^{i-}}{\mathcal{T}_q^{+-}} = \frac{\mathcal{T}_q^{i+}}{\mathcal{T}_q^{++}} \equiv v_\perp^i(x, \boldsymbol{k}_\perp).
    \label{eq:velocity_equivalence}
\end{equation}
This equality demonstrates that within on-shell constituent kinematics, the convective transport velocities of energy and momentum in the transverse plane are locally equal in $(x, \boldsymbol{k}_\perp)$, even in the presence of target polarization. In interacting field theory, dynamical gluon fields and strong interactions break this strict proportionality, rendering any discrepancy between energy and momentum transport a direct signature of off-shell QCD dynamics.

\subsection{From 2D transverse stress to 3D pressure anisotropy}
In Ref.~\cite{Lorce:2023zzg}, the transverse stress sub-tensor $\mathcal{T}_q^{ij}$ was decomposed into the two-dimensional (2D) isotropic transverse pressure $\sigma_q$, shear stress $\Pi_q$, polarized shear $\Pi_q^S$, and antisymmetric stress $\Pi_q^A$ (internal torque)~\cite{Polyakov:2018zvc,Lorce:2023zzg,Kaur:2026wxm} as
\begin{equation}
    \begin{aligned}
        \sigma_q & = \frac{1}{2}\Pi_q = \frac{1}{2P^+}\left[ \boldsymbol{k}_\perp^2 f^\perp + M \epsilon_T^{k_T S_T} f_T^- \right],                           \\
        \Pi_q^A  & = \frac{1}{2}\Pi_q^S = \frac{1}{2P^+}\left[ \lambda \boldsymbol{k}_\perp^2 f_L^\perp + M(\boldsymbol{k}_\perp\cdot\boldsymbol{S}_\perp) f_T^+ \right],
    \end{aligned}
\end{equation}
yielding the unpolarized isotropic transverse pressure $\sigma_{q,\text{unpol}}(x,\allowbreak \boldsymbol{k}_\perp^2) = \frac{\boldsymbol{k}_\perp^2}{2P^+} f^\perp(x, \boldsymbol{k}_\perp^2)$.

However, because the $\nu = -$ column was undetermined in general QCD, the longitudinal normal stress $\mathcal{T}_q^{--}$ remained inaccessible, leaving previous mechanical descriptions based on gravitational TMDs confined to the 2D transverse plane. On the null-plane $x^+ = \text{const}$, $x^-$ serves as the longitudinal spatial coordinate, and $\mathcal{T}_q^{--}$ represents the physical flux of minus-momentum across surfaces of constant $x^-$.

With $\mathcal{T}_q^{--}$ determined by the on-shell constituent condition (Eq.~\eqref{eq:Tmm}), we can directly compare the longitudinal normal stress with the transverse isotropic pressure. We define the dimensionless, boost-invariant longitudinal-to-transverse stress ratio $\mathcal{R}_{L/T}$ as
\begin{equation}
    \begin{aligned}
        \mathcal{R}_{L/T}(x, \boldsymbol{k}_\perp^2) & \equiv \frac{(P^+/M)^2\, \mathcal{T}_{q,\text{unpol}}^{--}(x, \boldsymbol{k}_\perp^2)}{\sigma_{q,\text{unpol}}(x, \boldsymbol{k}_\perp^2)} \\
        & = \left( \frac{m^2 + \boldsymbol{k}_\perp^2}{x \boldsymbol{k}_\perp^2} \right) \frac{f_3(x, \boldsymbol{k}_\perp^2)}{f^\perp(x, \boldsymbol{k}_\perp^2)}.
    \end{aligned}
    \label{eq:RLT}
\end{equation}
This ratio extends the 2D transverse stress picture to a 3D continuum mechanical description in momentum space, quantifying the local pressure anisotropy inside the hadron. By transverse rotational invariance, collinear partons carry no transverse momentum flux to exert transverse pressure, causing the transverse isotropic pressure $\sigma_{q,\text{unpol}} \propto \boldsymbol{k}_\perp^2$ to vanish as $\boldsymbol{k}_\perp \to 0$. In contrast, the longitudinal normal stress $\mathcal{T}_q^{--}(x, \boldsymbol{k}_\perp^2)$ remains strictly finite and positive at $\boldsymbol{k}_\perp = 0$, reflecting the non-zero longitudinal confinement tension carried by collinear partons. Consequently, $\mathcal{R}_{L/T}$ scales as $1/\boldsymbol{k}_\perp^2$ in the collinear limit $k_\perp \ll m$. This scaling reflects that stress transmission in the collinear limit is strictly longitudinal, whereas finite transverse motion is required to generate transverse pressure. While spatial mechanical observables such as the radial pressure $p(r)$, shear $s(r)$, and the von Laue stability condition require off-forward matrix elements with non-zero momentum transfer $\Delta$ and spatial Fourier transformation from gravitational form factors~\cite{Polyakov:2002yz,Burkert:2018bqq,Lorce:2018egm,Polyakov:2018zvc,Freese:2021czn}, the ratio $\mathcal{R}_{L/T}$ characterizes internal mechanical stress anisotropy directly in the forward parton momentum space probed by semi-inclusive deep-inelastic scattering.

\begin{figure*}[t]
    \centering
    \includegraphics[width=\textwidth]{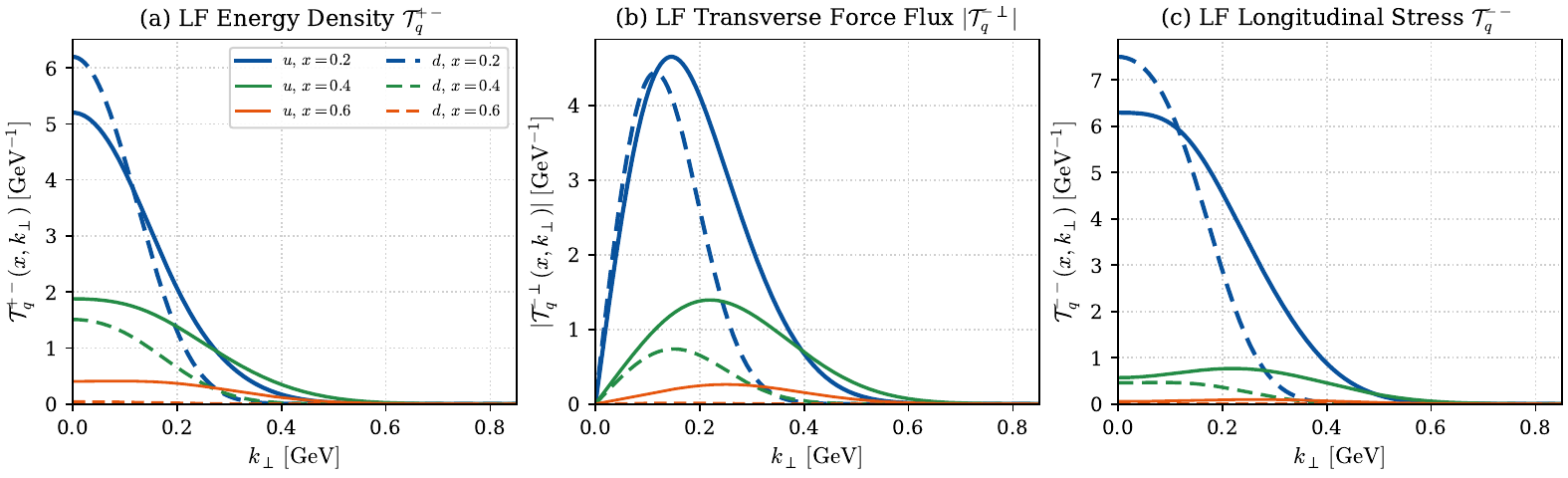}
    \caption{Unpolarized minus-sector energy-momentum tensor components in the proton evaluated at reference light-front (LF) momentum $P^+ = 1.0~\text{GeV}$. Panel (a) shows the LF energy density $\mathcal{T}_q^{+-}(x, k_\perp)$, panel (b) shows the transverse LF force flux magnitude $|\mathcal{T}_q^{-\perp}(x, k_\perp)| = \frac{M^2}{(P^+)^2} k_\perp f_3$, and panel (c) shows the longitudinal normal stress $\mathcal{T}_q^{--}(x, k_\perp)$, plotted as functions of $k_\perp$ for the $u$ quark (solid curves) and $d$ quark (dashed curves) at $x = 0.2, 0.4, 0.6$.}
    \label{fig:minus_components}
\end{figure*}

\section{Constituent quark model demonstration}
\label{sec:numerical}

To illustrate the physical behavior of these distributions, we implement a relativistic constituent quark-diquark model with scalar and axial-vector diquarks and Melosh spin rotations~\cite{Lu:2010dt}, building on the relativistic diquark framework~\cite{Jakob:1997wg,Gamberg:2007wm,Bacchetta:2008af,Gurjar:2022rcl,Sharma:2024lal} with one-gluon exchange final-state interactions~\cite{Brodsky:2002rv,Lu:2004au} to generate the complete set of T-odd distributions, which also satisfy characteristic model relations among TMDs and GPDs~\cite{Meissner:2007rx,Lorce:2014hxa}. For the proton, model parameters~\cite{Lu:2010dt} are constituent quark mass $m = 0.22~\text{GeV}$, proton mass $M = 0.938~\text{GeV}$, scalar diquark mass $M_s = 0.50~\text{GeV}$, axial-vector diquark mass $M_a = 0.70~\text{GeV}$, and harmonic oscillator momentum wave-function scales $\beta_s = 0.25~\text{GeV}$ and $\beta_a = 0.17~\text{GeV}$. Calculations are evaluated in the standard reference frame $P^+ = 1.0~\text{GeV}$.

In Fig.~\ref{fig:minus_components}, we present the three unpolarized light-front EMT components containing the minus index, namely the partonic energy density $\mathcal{T}_q^{+-}$, the transverse force flux magnitude $|\mathcal{T}_q^{-\perp}| \equiv |\mathcal{T}_{q,\text{unpol}}^{-i}| = \frac{M^2}{(P^+)^2} k_\perp f_3$, and the longitudinal normal stress $\mathcal{T}_q^{--}$, evaluated across $k_\perp \in [0, 0.85]~\text{GeV}$ for $x = 0.2, 0.4, 0.6$. Because the unpolarized twist-2 and twist-4 TMDs $f_1$ and $f_3$ represent positive-definite probability distributions in constituent frameworks, all three quantities are positive semi-definite throughout the physical phase space. Panels (a) and (c) show that the energy density $\mathcal{T}_q^{+-}$ and longitudinal normal stress $\mathcal{T}_q^{--}$ attain non-zero finite values at $k_\perp = 0$, dictated by $(m^2/(2xP^+)) f_1(x, 0)$ and $(M^2/(P^+)^2)(m^2/(2xP^+)) f_3(x, 0)$, respectively, reflecting the finite energy and longitudinal confinement tension carried by strictly collinear partons. In contrast, panel (b) shows that the transverse force flux magnitude $|\mathcal{T}_q^{-\perp}|$ vanishes linearly as $k_\perp \to 0$, as required by transverse rotational invariance, and peaks at the constituent Fermi scale $k_\perp \approx 0.15$-$0.20~\text{GeV}$. Flavor-wise, the curves reflect the diquark spin and wave-function structures. The $d$ quark couples solely to the axial-vector diquark $A$, whose narrower wave-function scale $\beta_a = 0.17~\text{GeV}$ concentrates its probability density near $k_\perp \sim 0$, causing both the energy density $\mathcal{T}_d^{+-}$ and longitudinal normal stress $\mathcal{T}_d^{--}$ at $x = 0.2$ to start slightly above their $u$-quark counterparts at $k_\perp = 0$. However, this narrower momentum width causes the $d$-quark distributions to fall off far more rapidly with increasing $k_\perp$. At larger longitudinal momentum fractions $x = 0.4$ and $0.6$, the $u$ quark heavily dominates across all $k_\perp$, exceeding the $d$ quark by an order of magnitude at $x = 0.6$.

\begin{figure*}[t]
    \centering
    \includegraphics[width=0.85\textwidth]{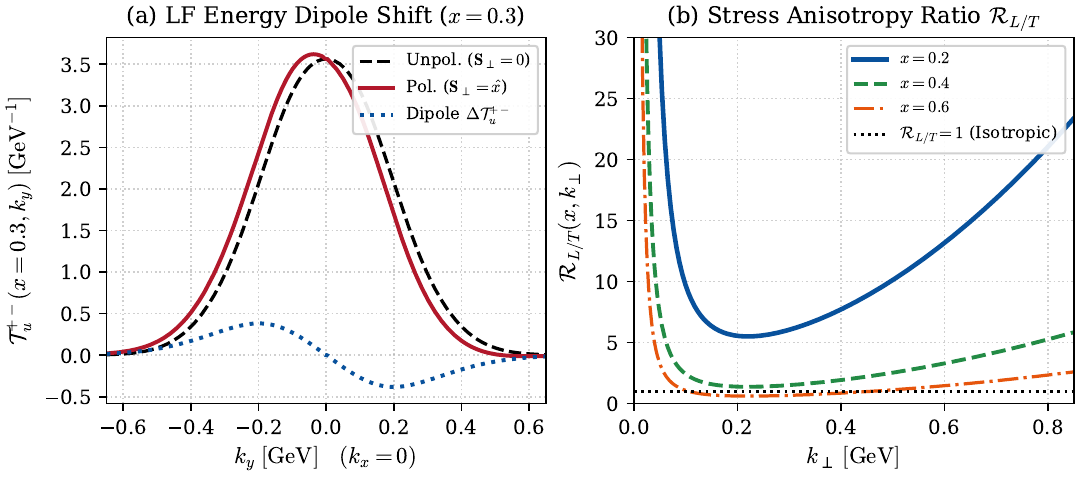}
    \caption{Sivers lateral shift and 3D mechanical stress anisotropy evaluated at reference scale $P^+ = 1.0~\text{GeV}$. Panel (a) illustrates the transverse spin distortion of the light-front energy density along $k_y$ at $k_x = 0$ for $x = 0.3$ in a proton polarized along $\boldsymbol{S}_\perp = \hat{x}$, showing the unpolarized profile, the polarized profile, and the dipole shift $\Delta \mathcal{T}_u^{+-}$ generated by one-gluon exchange final-state interactions. Panel (b) shows the boost-invariant mechanical stress anisotropy ratio $\mathcal{R}_{L/T}(x, k_\perp)$ at $x = 0.2, 0.4, 0.6$, with the dotted horizontal line indicating isotropic stress balance ($\mathcal{R}_{L/T} = 1$).}
    \label{fig:stress_anisotropy}
\end{figure*}

Figure~\ref{fig:stress_anisotropy} highlights the polarized and anisotropic continuum mechanical features revealed by the completed EMT. Figure~\ref{fig:stress_anisotropy}(a) demonstrates the effect of transverse target polarization on the light-front energy density along the $k_y$ axis at $k_x = 0$ for $x = 0.3$, for a proton polarized along $\boldsymbol{S}_\perp = \hat{x}$. In this orientation, one-gluon exchange final-state interactions generate an attractive chromodynamic lensing force, yielding a negative $u$-quark Sivers function $f_{1T}^{\perp u} < 0$. The resulting dipole distortion $\Delta \mathcal{T}_u^{+-} \propto k_y f_{1T}^\perp$ is odd in $k_y$. When superimposed on the symmetric unpolarized profile $\mathcal{T}_{u,\text{unpol}}^{+-}$, this dipole shift visibly displaces the peak of the physical energy density $\mathcal{T}_u^{+-}(\boldsymbol{S}_\perp)$ toward negative $k_y$, directly establishing that chromodynamic lensing displaces not only the quark number density, but also the internal energy density of the nucleon. Figure~\ref{fig:stress_anisotropy}(b) displays the boost-invariant mechanical stress ratio $\mathcal{R}_{L/T}(x, k_\perp)$. As anticipated from the kinematic scaling in Eq.~\eqref{eq:RLT}, the ratio diverges as $1/k_\perp^2$ in the collinear limit $k_\perp \to 0$ where internal stress transmission is exclusively longitudinal. As $k_\perp$ increases toward the constituent scale $k_\perp \sim 0.3$-$0.5~\text{GeV}$, intrinsic transverse Fermi motion builds up the transverse pressure $\sigma_q$, driving the ratio smoothly downward toward the isotropic mechanical balance benchmark $\mathcal{R}_{L/T} \sim 1$ indicated by the dotted horizontal line. The transition occurs at progressively larger $k_\perp$ as $x$ increases, reflecting the higher longitudinal momentum carried by large-$x$ valence quarks.

\section{Conclusion}
\label{sec:conclusion}

In summary, we have shown that within relativistic light-front constituent frameworks, treating the active quark as an effective on-shell quasiparticle provides a consistent and closed scheme to determine the $\nu = -$ sector of the momentum-space energy-momentum tensor (EMT). This matching expresses all 32 gravitational transverse-momentum distributions (TMDs) in terms of canonical vector TMDs up to twist-4, accommodating target polarization and T-odd effects from one-gluon exchange final-state interactions.

The completed energy density $\mathcal{T}^{+-}_q$ provides a direct momentum-space representation of how light-front energy $k^-$ is partitioned among partons. For a transversely polarized target, the Sivers effect generates a lateral dipole shift in $\mathcal{T}^{+-}_q$, demonstrating that Burkardt's chromodynamic lensing mechanism displaces not only the quark number density, but also the internal energy density of the nucleon. Furthermore, the convective transport velocities of light-front energy and inertia prove locally equal in $(x, \boldsymbol{k}_\perp)$, reflecting an intrinsic kinematic property of factorized on-shell descriptions that provides a baseline to quantify dynamical off-shell corrections in interacting field theories.

Access to the longitudinal normal stress $\mathcal{T}_q^{--}$ elevates the mechanical description of the nucleon beyond the two-dimensional transverse plane into a three-dimensional momentum-space continuum. Comparing longitudinal and transverse stresses through the boost-invariant ratio $\mathcal{R}_{L/T}$ charts the mechanical transition from longitudinal confinement stress dominance with $\mathcal{R}_{L/T} \gg 1$ at low $k_\perp$ to transverse isotropic stress balance $\mathcal{R}_{L/T} \sim 1$ at intermediate $k_\perp$ in the valence region. Complementing these general features, numerical calculations within a relativistic quark-diquark framework reveal distinct flavor profiles for $u$ and $d$ quarks stemming from diquark spin and wave-function scales.

We note that the present determination of the $\nu = -$ sector is based on the on-shell dispersion $k^- = (m^2 + \boldsymbol{k}_\perp^2)/(2x P^+)$, isolating the kinetic energy and kinetic stress carried by the bound constituents. In full interacting field theory, the light-front Hamiltonian also incorporates dynamical interaction contributions from gauge-field constraints and higher-twist quark-gluon correlators. Extending this momentum-space framework to include dynamical interaction kernels will provide valuable insight into how binding forces and off-shell dynamics shape the complete hadronic EMT. These constituent model insights offer concrete benchmark expectations for ongoing and future extractions of hadronic continuum mechanics from lattice quantum chromodynamics and at the Electron-Ion Collider.

\section*{Acknowledgements}
H.D. would like to thank the Science and Engineering Research Board, Anusandhan-National Research Foundation, Government of India under the scheme SERB-POWER Fellowship (ref.~no.~SPF/2023/000116) for financial support.

\bibliographystyle{elsarticle-num}
\bibliography{refs}

\end{document}